# Swin UNETR segmentation with automated geometry filtering for biomechanical modeling of knee joint cartilage

Reza Kakavand[1], Peyman Tahghighi[1], Reza Ahmadi[1], W. Brent Edwards[1,2,3], Amin Komeili[1,2,3*]

[1]Department of Biomedical Engineering, Schulich School of Engineering, University of Calgary

[2]McCaig Institute for Bone and Joint Health, University of Calgary, Calgary, Canada.

[3]Human Performance Laboratory, Faculty of Kinesiology, University of Calgary, Calgary, Canada

**Corresponding author: Amin Komeili**

\* <amin.komeili@ucalgary.ca>

**Address: CCIT216, 2500 University Drive NW, Calgary, AB, T2N 1N4**

# Abstract

Simulation studies, such as finite element (FE) modeling, offer insights into knee joint biomechanics, which may not be achieved through experimental methods without direct involvement of patients. While generic FE models have been used to predict tissue biomechanics, they overlook variations in population-specific geometry, loading, and material properties. In contrast, subject-specific models account for these factors, delivering enhanced predictive precision but requiring significant effort and time for development. This study aimed to facilitate subject-specific knee joint FE modeling by integrating an automated cartilage segmentation algorithm using a 3D Swin UNETR. This algorithm provided initial segmentation of knee cartilage, followed by automated geometry filtering to refine surface roughness and continuity. In addition to the standard metrics of image segmentation performance, such as Dice similarity coefficient (DSC) and Hausdorff distance, the method's effectiveness was also assessed in FE simulation. Nine pairs of knee cartilage FE models, using manual and automated segmentation methods, were developed to compare the predicted stress and strain responses during gait. The automated segmentation achieved high Dice similarity coefficients of 89.4% for femoral and 85.1% for tibial cartilage, with a Hausdorff distance of 2.3 mm between the automated and manual segmentation. Mechanical results including maximum principal stress and strain, fluid pressure, fibril strain, and contact area showed no significant differences between the manual and automated FE models. These findings demonstrate the effectiveness of the proposed automated segmentation method in creating accurate knee joint FE models. The automated models developed in this study have been made publicly accessible to support biomechanical modeling and medical image segmentation studies (https://data.mendeley.com/datasets/dc832g7j5m/1).

# 1 Introduction

Joint and soft tissue simulation methods, particularly using finite element (FE) modeling, have offered significant insights into the biomechanics of the knee joint, eliminating the need for direct experimentation on patients (Halilaj et al., 2018; Kakavand, Palizi, et al., 2023; Kakavand, Rasoulian, et al., 2023; Klets et al., 2018; Mononen et al., 2018; Orozco et al., 2018; Park et al., 2019; Shu et al., 2018; Thomas et al., 2020). Generic models, based on average biomechanical behavior (Ahmadi et al., 2024; Cooper et al., 2019; Komeili et al., 2019), have been useful but fail to consider individual variations. Hence, the shift towards personalized subject-specific models (Bolcos et al., 2022; Chokhandre et al., 2023; Henak et al., 2013, 2014; Lavikainen et al., 2023; Li et al., 2023) is inevitable for more accurate predictions of biomechanical response of the tissue. Computed tomography (CT) and magnetic resonance images (MRIs) serve as preferred medical imaging techniques for capturing subject-specific geometry of knee joints. However, the manual construction of these models is time-consuming (Liukkonen et al., 2017; Myller et al., 2020) and lacks reproducibility (Erdemir et al., 2019; Kang et al., 2015; Koo et al., 2005). To expedite this process, convolutional neural networks (CNNs) have demonstrated effective performance in knee tissue segmentation (Norman et al., 2018; Prasoon et al., 2013). For instance, Ambellan et al. (2019) utilized CNN for 3D knee cartilage segmentation and achieved Dice similarity coefficient (DSC) values of 89.9% and 85.6% for femoral cartilage (FC) and tibial cartilage (TC), respectively. Despite advancements, reproducible and applicable segmentation methods tailored to computational modeling are still needed.

Amongst CNNs models, U-Net based models have shown exceptional performance in segmenting medical images (Burton II et al., 2020; Ronneberger et al., 2015), but they may struggle with capturing the influence of pixels or regions that are spatially distant from each other (Hatamizadeh et al., 2022). UNETR is a neural network architecture that combines the strengths of UNet and transformer models for accurate image segmentation tasks (Hatamizadeh et al., 2022). For instance, Swin UNETR (Hatamizadeh et al., 2021) leverages Swin transformers and a modified Vision Transformer (ViT), achieving superior accuracy and efficiency in various benchmarks (Çiçek et al., 2016; Dosovitskiy et al., 2020).

Despite advancements in image segmentation, manual correction remains necessary for tasks like surface smoothness enhancement and abnormal morphology correction for FE modeling. Statistical shape modeling (SSM) (Cootes et al., 1995) was used for geometry adjustment and filling holes (Ambellan et al., 2019; Kakavand et al., 2024). However, in the context of cartilage segmentation, the presence of holes and fissures could signal osteoarthritis (OA), and employing SSM models might mistakenly fill these features, consequently restricting their effectiveness to healthy cartilage exclusively. Moreover, a gap exists between the contemporary advanced automated segmentation models (Ambellan et al., 2019; Norman et al., 2018; Prasoon et al., 2013) and their utilization within biomechanical modeling. This creates a lack of readily available automated segmentation models suitable for subject-specific biomechanical modeling of knee cartilage in the public domain.

The primary goal of this research was to develop an automated approach for constructing knee joint cartilage geometry of healthy and OA cartilage for FE modeling, aiming to achieve comparable accuracy to the manual segmentation method. The objectives of the present study include 1) training a 3D Swin UNETR transformer coupled with automated filtering of knee joint cartilage, 2) computational modeling of the knee joint cartilage, and 3) evaluating the accuracy of mechanical responses predicted by the automated segmentation method against manually segmented models.

## 2 Method

### 2.1 Data

MRIs from 507 individuals (aged 61.9 ± 9.3 years; BMI 29.27 ± 4.52 kg/m²; image resolution of 0.36×0.36×0.7 mm) were obtained from the Osteoarthritis Initiative (OAI) database (https://nda.nih.gov/oai/), consisting of 262 males and 245 females. Manual segmentation of this dataset was performed by skilled users from the Zuse Institute Berlin (Ambellan et al., 2019). The dataset covered all grades of OA but predominantly included severe cases, with 60 MRIs graded as Kellgren-Lawrence (KL) 0, 77 as grade 1, 61 as grade 2, 151 as grade 3, and 158 as grade 4 OA.

To assess the effectiveness of the Swin UNETR model, a random 5-fold cross-validation method was employed, with each fold comprising 405 MRIs for training and 102 MRIs for testing (*Figure 1.SM in the supplementary material*). Regarding the evaluation of FE models, nine samples were randomly selected from the test set due to the extremely time-consuming nature of FE modeling of 102 samples, which currently lacks automation in steps such as meshing, material property assignment, and loading.

## 2.2 Swin UNETR

A 3D Swin UNETR with a patch size of 2, a window size of 7, and an initial feature size of 48 were utilized. The four output features obtained from the Swin transformer (shown by red arrows) were inputted into 3D U-Net and convolutional blocks for reconstructing an image of the same size as the input (Figure 1). The DSC and focal loss were employed as evaluation metrics. Our proposed method was implemented in Python using the Pytorch library. All models were trained using a batch size of 8, with the Adam optimizer (Kinga et al., 2015; Paszke et al., 2017) and an initial learning rate of 0.0001. To prevent overfitting, an early stopping technique was implemented during the training. The proposed model was evaluated using 5-fold cross-validation. Throughout the training process, each MRI was resized to 128 × 128 × 160 (to reduce the computational costs) and randomly cropped to a size of 96 × 96 × 96. Additionally, window center adjustment was performed on the MRIs as a preprocessing step to translate them to a range of [0, 1].

The data augmentation and Swin UNETR implementation were done using the MONAI library (https://docs.monai.io/en/stable/). All the other variables were the default settings in the MONAI. The Swin UNETR models, codes and a video demonstrating the implementation of the model were made publicly available (https://data.mendeley.com/datasets/dc832g7j5m/1).

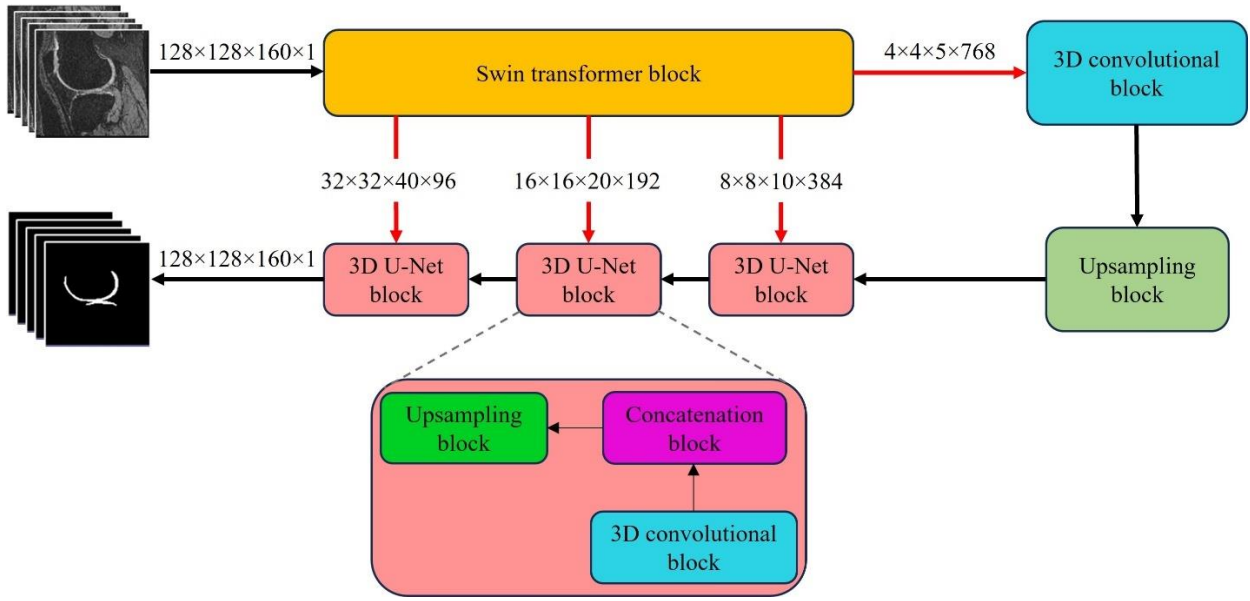

Figure 1. The structure of the employed Swin UNETR segmentation model. It was a combination of a Swin transformer as the encoder and 3D U-Net blocks as the decoder. Swin transformer extracted hierarchical representation from a given MRI and 3D U-Nets utilized these representations to construct the segmentation mask.

## 2.3 Automated filtering

The output of the Swin UNETR, i.e. automated segmentation model, underwent an automated filtering process to smooth the cartilage surface geometry. This process included mesh repairing, point sampling, smoothing, and rigid registration and was implemented in a Python script that is publicly available at (https://data.mendeley.com/datasets/dc832g7j5m/1). Mesh repairing (Muntoni & Cignoni, 2021) consisted of merging close vertices, merging close per-wedge texture coordinates, removing duplicate faces and vertices, removing folded and null faces, removing floating faces, removing t-vertices and unreferenced vertices, and removing non-manifold edges and vertices *whose detailed definition is provided in the supplementary material.* Montecarlo (Covre et al., 2022) and Poisson-disk sampling (Corsini et al., 2012; Rooks et al., 2021) were used to achieve uniformly distributed points. The resulting mesh was then smoothed using a Laplacian method (Sorkine, 2005) which moves each vertex in the average position of the neighboring vertices only if the new position still lies on the original surface. Lastly, the Coherent Point Drift (CPD) method (Myronenko & Song, 2010) was used for the rigid registration (Figure 2). CPD iteratively applies rotation, scaling, and translation to align the smoothed geometry with the Swin UNETR outcome (Figure 2). Before using the automated filtering method, the output of Swin UNETR, which is an image, was manually converted to a CAD format (for example, using ITK-SNAP software). A tutorial video was recorded about converting the segmented images to a CAD, which is available at https://data.mendeley.com/datasets/dc832g7j5m/1.

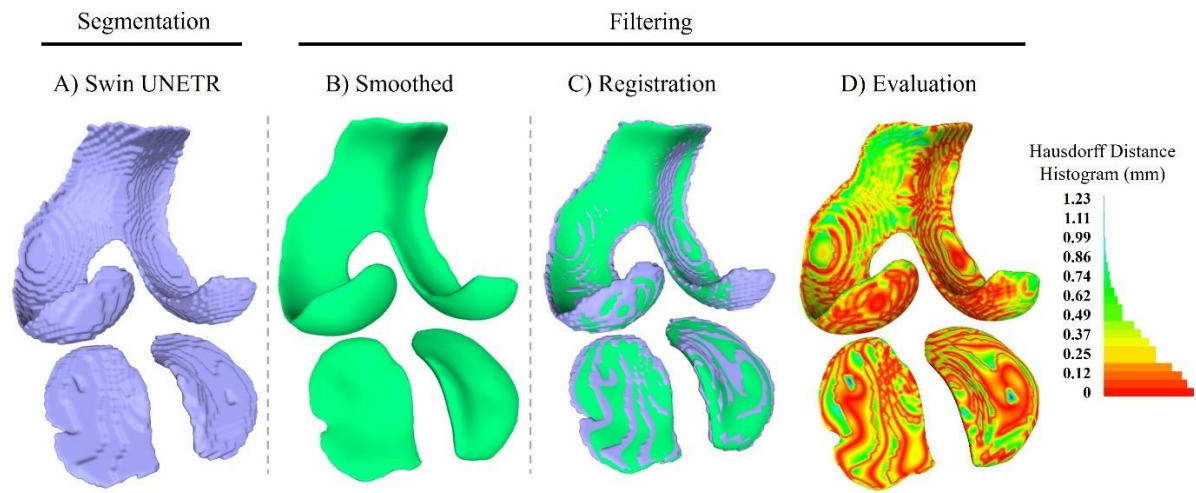

Figure 2. The workflow of automated filtering. A) the outcome of Swin UNETR segmentation model. B) mesh repairing, point sampling, smoothing was applied to the Swin UNETR outcome. C) the smoothed geometry was fitted to the Swin UNETR outcome using rigid registration. D) Hausdorff distance histogram for evaluating the performance of the segmentation (please refer to Section 2.4 for the evaluation metrics used in this study). The automated filtering procedure was implemented in a custom Python script, publicly available at (https://data.mendeley.com/datasets/dc832g7j5m/1).

## 2.4 Evaluation metrics

The metrics to evaluate the segmentation performance of FC and TC using the Swin UNETR and filtering methods included DSC, Hausdorff distance, average distance, and the percentage of surface area associated with a distance greater than 1 mm between the two methods. The DSC measures the overlap between the segmented regions and the manual segmentation as the ground truth (intersection over union). The Hausdorff distance quantifies the maximum distance from the nearest neighbor (Huttenlocher et al., 1993) between corresponding points on the segmented surface and the ground truth. The other calculated parameter to assess the accuracy of the automated method was the average distance, which represents the average separation between the segmented surface and the ground truth surface. Lastly, the metric ∆area%>1mm quantifies the percentage of surface area where the distance between the segmented regions and the ground truth exceeds 1 mm.

In addition to image segmentation metrics, the effectiveness of the automated model in predicting mechanical responses of the tissue was evaluated through FE simulation. This approach was employed to examine the impact of errors in automatic segmentation on the prediction of tissue stresses and strains. The cartilage from nine knee MRIs

(age: 58±8.5 years; BMI: 28.3±3.6 kg/m$^2$; 2 MRIs graded as KL 0, 1 as grade 1, 2 as grade 2, 2 as grade 3, and 2 as grade 4 OA) underwent segmentation through both manual and automated methods, resulting in a combined total of 18 FE models (nine manual and nine automated). The automated segmentation process was described in section 2.2. The automated filtering was performed for both manual and automated models (Figure 3). Finally, the mechanical responses of the manual and automated FE models were compared considering a loading condition that simulates gait.

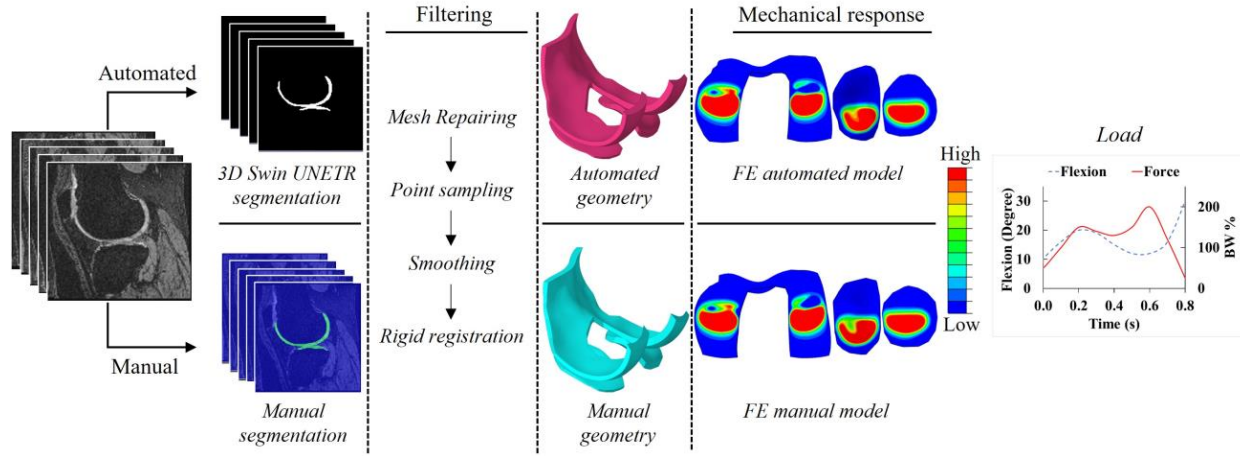

Figure 3. FE model development workflow used in this study. Automated and manual segmentations were developed from knee MRIs in OAI database ( https://nda.nih.gov/oai/ ). An automated filtering technique was applied to the geometry outcome of 3D Swin UNETR, giving smooth surfaces suitable for mesh generation for FE analysis (Section 2.3). Gait loading was applied to both manual and automated FE models, and their mechanical responses were compared, including cartilage contact mechanics and pore pressure.

## 2.5 Computational modeling

Material and FE modeling

A biphasic constitutive law proposed by Federico and Gasser (2010) and Federico and Grillo (2012) was used for cartilage mechanical behavior. This constitutive law includes an incompressible fluid phase and a fibril-reinforced solid/matrix phase. The collagen fibrils were divided into isotropic and anisotropic components, assuming that the matrix was isotropic and inhomogeneous along its thickness. The anisotropic fibrils were responsible for the directional orientation of the fibrillar network (Komeili et al., 2020).

The material constants for the biphasic constitutive law can be found in ***Table 1.SM in the supplementary material***. To determine the material properties of the extracellular matrix (ECM), the results of creep indentation experiments conducted by Pajerski (2010) and Athanasiou et al. (1991) on human knee cartilage were used. For the material characteristics of collagen fibrils, the information provided by Komeili et al. (2020) was used. For a more detailed explanation of the cartilage constitutive laws and their material constants used to describe cartilage behavior, readers can find *the supplementary material*. In essence, the state of stress was defined by the following:

$$\sigma = -pI + \emptyset_0 \sigma_0 + \emptyset_1 (\sigma_{1i} + \sigma_{1a}) \qquad \text{eq. 1}$$

Where $\sigma$ is the total stress in the tissue, $p$ is the hydrostatic interstitial fluid pressure, $I$ is the unity tensor, and $\emptyset$ is the volume fraction. Subscripts 0 and 1 denote matrix and collagen fibrils, respectively. The matrix was considered isotropic, while the collagen fibrils were divided into two parts of isotropic ($\sigma_{1i}$) and anisotropic ($\sigma_{1a}$).

The knee cartilage mesh was defined using hexahedral pore pressure elements (C3D8P) (*Figure 2.SM in the supplementary material* depicts the mesh processing in detail). A surface-to-surface contact with frictionless tangential behavior was implemented to simulate the contact mechanics between the cartilage surfaces. The ligaments insertion points were determined from MRIs using ITK-SNAP software and confirmed by an experienced musculoskeletal radiologist. The Anterior cruciate ligament (ACL), posterior cruciate ligament (PCL), medial and lateral collateral ligaments (MCL, LCL) were considered as bi-linear springs, capable of withstanding tension but not compression. The tensile strength for the ACL was set at k=380 N/mm (Donahue et al. (2002)), while k=200 N/mm was used for the PCL (Momersteeg et al. (1995)). For MCL and LCL, the tensile stiffness was set at k=100 N/mm, with data from Bolcos et al. (2018) and Momersteeg et al. (1995).

The middle-central position between the medial and lateral epicondyles of the femur was used as the reference point for coupling the femur surface to the loading (Mononen et al., 2016, 2023). The bottom nodes of TC were fixed. The cartilage surfaces at the calcified zone were assumed impermeable, while the pore pressure of the articular cartilage surfaces was set to zero, permitting free fluid flow. The stance phase of gait was simulated by applying a combination of an indentation load and a flexion angle at the reference point (Mohammadi et al., 2020; Mononen et al., 2016) (Figure 3). A settling step was considered before the stance phase, where a load of 30 N was applied for one second on the reference point of the femur to make the initial contact of cartilage surfaces. Abaqus/CAE software 2018 (Dassault Systems Simulia Corp., Johnston, RI, USA) was used for the FE modeling. The FE mesh was generated in HyperMesh 2019 (Altair Inc, Santa Ana, CA).

## 2.6 Statistical analysis

To assess and compare the mechanical response between the manual and automated models, five parameters were used throughout the stance simulation: maximum principal stress, fluid pressure, maximum principal strain, fibril strain, and contact area. The comparison of the first four parameters was conducted for both the superficial and deep zones, whereas the contact area was only evaluated on the articular surface of the cartilage. For each zone, the average and peak values of these parameters were analyzed and compared.

To make multiple comparisons, statistical parametric mapping (SPM) was used. This method was chosen due to its inherent advantage in accommodating multiple comparisons through time, unlike traditional 0-D approaches such as the parametric t-test (Jahangir et al., 2022). The SPM t-test was employed for paired two independent samples, with a confidence interval of 0.05. For SPM implementation, a Python package was used available at https://spm1d.org/#.

## 3 Results

Table 1 provides the evaluation metrics for the Swin UNETR segmentation before and after the filtering process (Figure 3). The output of the Swin UNETR method resulted in DSCs of [89.3, 85.5] %, and Hausdorff distance of [2.51, 2.39] mm for [FC, TC], respectively. The corresponding values after the filtering process were [89.4, 85.1] % and [2.35, 2.51] mm for [FC, TC], respectively. Regarding the average distance, the Swin UNETR method and automated filtering yielded values of [0.20, 0.19] mm and [0.23, 0.17] mm for [FC, TC], respectively. Additionally, FC and TC segmentations using the Swin UNETR method exhibited a Δarea% > 1mm of [2.12, 2.24] %, respectively. Following the automated filtering, these values were [2.31, 2.54] %, respectively. ***Figure 3.SM in the supplementary material*** provides a visual comparison between manual and Swin UNETR segmentations. The Swin UNETR segmentation process required approximately 5 minutes to generate FC and TC geometries compared to the approximately 6 hours of manual segmentation by an expert (approximately 5 hours for manual segmentation in ITK-SNAP software and about 1 hour for model smoothing in MeshLab).

Statistical analysis revealed no significant difference between the manual and automated FE models across all nine samples. Figure 4 illustrates SPM of maximum principal stress and strain, fluid pressure, fibril strain, and contact area over time. The values of all parameters remained within the critical thresholds, showing no significant difference (p-value > 0.05).

Figure 5 illustrates the distribution of mechanical responses across the surface and depth-wise at 20% and 80% of the stance phase for subject 1. Both the manual and automated FE models exhibited comparable distribution of parameters. *Figures 4.SM -7.SM* illustrate the distribution of each mechanical response for all cartilage samples.

The average and peak values of the mechanical parameters in the superficial and deep zones are illustrated in Figure 6. The dotted line represents the absolute difference between the two FE models. The contact region over the articular surface was projected into the deep zone to measure the mechanical parameters of this zone. The fluid pressure illustrated the largest error of 0.017 MPa. In the *supplementary material, Figures 8.SM – 11.SM* were plotted for each sample separately to provide a more detailed comparison of mechanical parameters between the automated and manual FE models. All these figures indicated no significant difference in the mechanical response obtained from the automated FE model compared to the manual one.

Table 1. DSC, Hausdorff distance (mm), average distance (mm), and percentage of surface area ($\Delta area\%$) associated with a distance greater than 1 mm for the output of the Swin UNETR before and after filtering compared to manual segmentation.

|  | DSC | Hausdorff distance (*mm*) | Average distance (*mm*) | $\Delta area\%$ > $1mm$ |
|---|---|---|---|---|
| FC Swin UNETR | 89.3±3.3 | 2.51±0.94 | 0.20±0.04 | 2.12±1.04 |
| TC Swin UNETR | 85.5±2.5 | 2.39±0.82 | 0.19±0.04 | 2.24±1.03 |
| FC filtered | 89.4±2.7 | 2.35±1.21 | 0.23±0.02 | 2.31±1.22 |
| TC filtered | 85.1±4.6 | 2.51±1.07 | 0.17±0.05 | 2.54±1.23 |

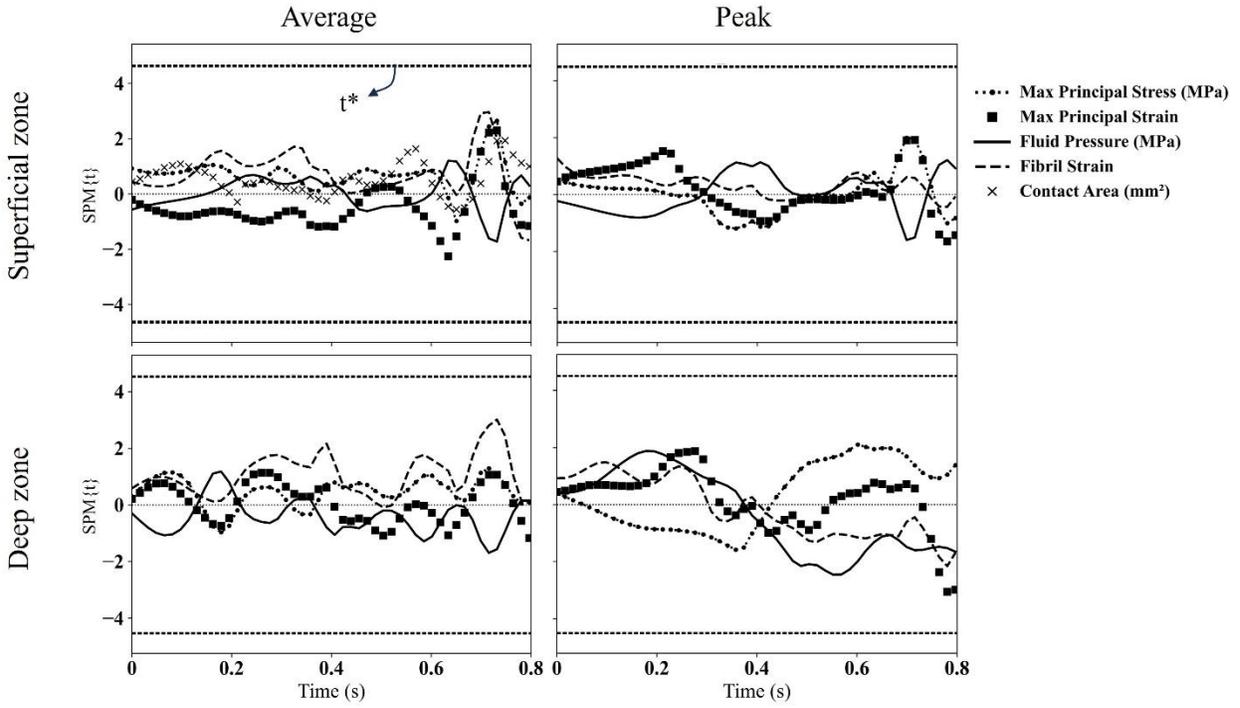

Figure 4. Statistical parametric mapping (SPM) as a function of time for 5 mechanical parameters in superficial and deep zones. The dashed line shows the t-critical corresponding to a p-value of 0.05. The Average column shows the average of the respected parameter over the contact region of all samples. Similarly, the Peak column shows the maximum value of the 5 parameters. The contact region over the articular surface was projected into the deep zone to calculate the parameters of the deep zone.

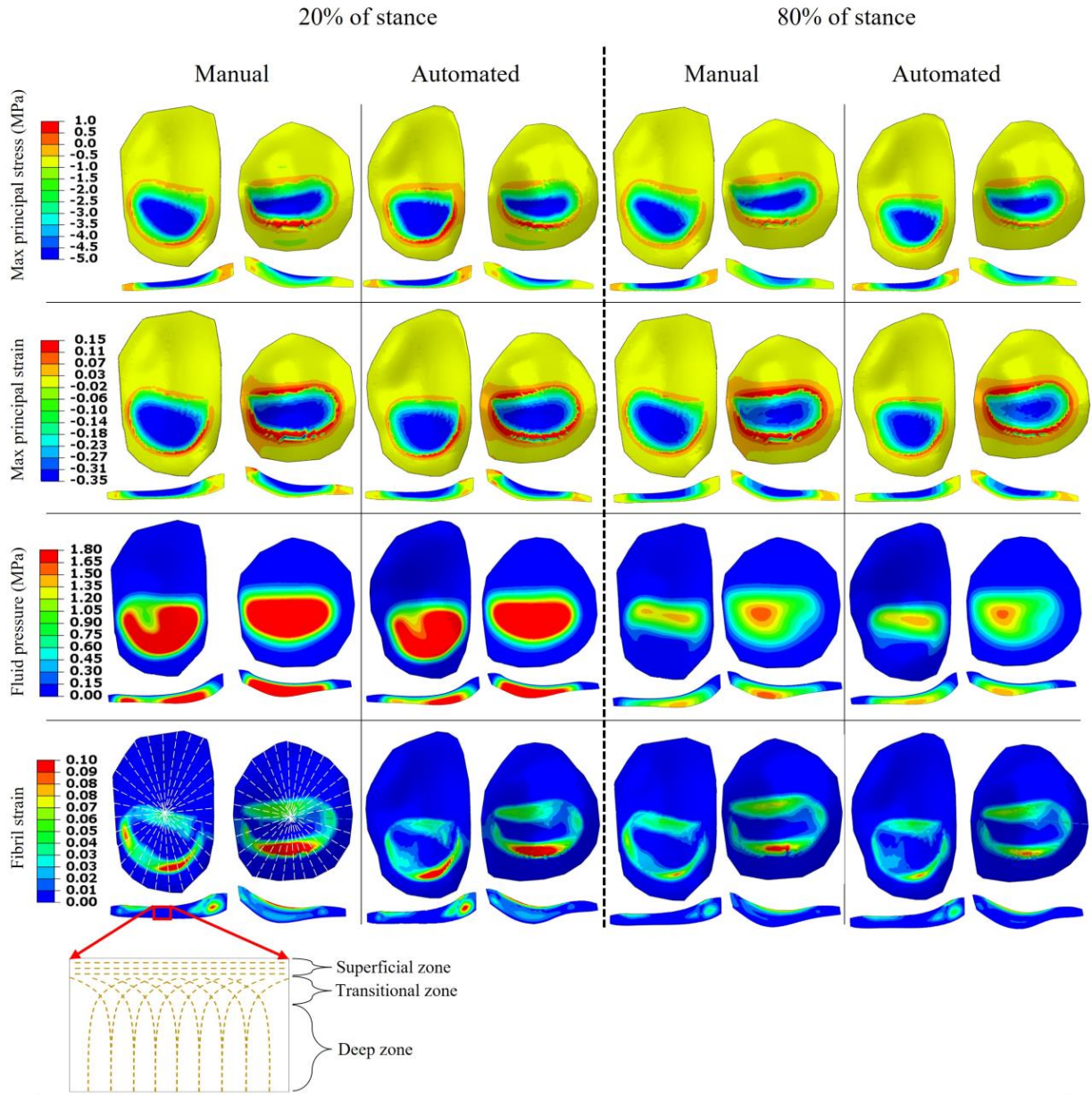

Figure 5. The distribution of maximum principal stress and strain, fluid pressure, and fibril strain over the surface and thickness of TC at 20% and 80% of the stance phase for subject 1. Fibril strain was measured along the fibril, and its distribution is depicted at 20 % of stance on the manual model by dashed lines (Kakavand, Rasoulian, et al., 2023). The depth-wise illustration was from the cross-section where the peak value occurred.

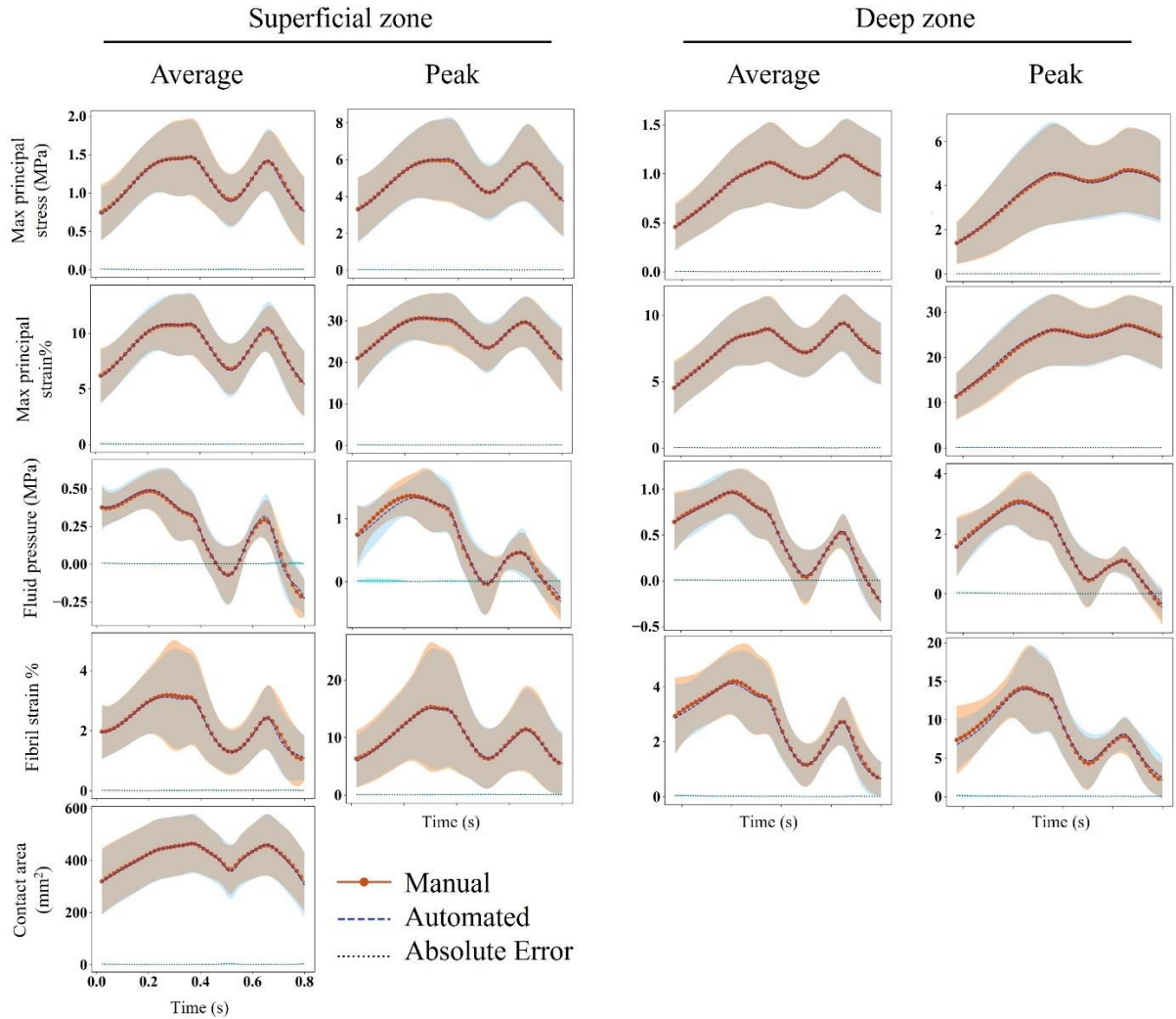

Figure 6. The average (over contact area) and maximum values of maximum principal stress and strain, fluid pressure, and fibril strain in the superficial and deep zones. The shaded region represents one standard deviation. The solid line with a circular marker represents the manual FE model, whereas the dashed line represents the FE model created from automated cartilage segmentation. The dotted line is the absolute difference between the two models. The contact region over the articular surface of the cartilage was projected into the deep zone to calculate the parameters of this zone.

# 4 Discussion

In the present study, an automated filtering method was applied to the outcome of a Swin UNETR segmentation model to produce personalized geometry from MRIs. The filtering process automated the post-processing operations associated with mesh repairing and smoothing, which are essential steps to increase the convergency rate in FE simulations (Figure 2) (Ambellan et al., 2019; Bruce et al., 2021; Clouthier et al., 2019). The results highlighted the success of the suggested automated segmentation approach, combining the Swin UNETR segmentation model with filtering, in producing precise knee joint FE models of knee cartilage.

Previous studies (Ambellan et al., 2019; Kakavand et al., 2024) evaluating knee segmentation utilized SSM (Cootes et al., 1995) on neural network segmentations of femur and tibia to address hole filling and surface smoothing. However, when dealing with cartilage segmentation, holes and irregular surface topology may represent fissures or damages in the surface of cartilage as a result of OA, which could be of significant interest. On the other hand, discretizing the cartilage surfaces for FE analysis directly after MRI segmentation without subsequent smoothing, presents a formidable challenge. This process, if feasible, can take hours or even days to complete. Typically, a trial-and-error approach is required to strike an appropriate balance between preserving surface properties and achieving a suitable mesh for FE studies. Utilizing SSM models to automate this process may inaccurately fill such holes, thereby limiting their applicability to healthy cartilage only. In the current research, a sequence of filtering processes (Section 2.3) was designed to preserve holes and damaged sites in the model (Figure 7). The Laplacian smoothing method (Laplacian smoothing surface preserving function in PyMeshLab) (Nealen et al., 2005; Sorkine, 2005) employed in the automated filtering performs a robust iterative smoothing while retaining geometric details. It relocates each vertex to the average position of its neighboring vertices, ensuring the new position remains on the original surface. Subsequently, the rigid registration method was employed to iteratively apply rotation, scaling, and translation to map the smoothed geometry onto the Swin UNETR outcome. This registration procedure aims to offset the potential volumetric shrinking effect of the smoothing while maintaining surface quality for FE modeling (Figure 7).

The average DSCs of the Swin UNETR models were 89.4% and 85.5% for FC and TC, respectively (Table 1). Applying the filters smoothed the cartilage surfaces but did not affect the DSCs significantly. The most impact of applied filters was on the average distance metric for FC, where it changed from 0.20 mm to 0.23 mm, a 15% change, after applying the smoothing filters (Table 1). However, this did not change the stress and strain distributions significantly (Figure 4). The metrics in Table 1 demonstrate a comparable performance compared to the other studies. For instance, Ambellan et al., (2019) reported DSC of 89.9% and 85.6% for FC and TC, respectively. Norman et al., (2018) reported DSC in the range of 77% to 88% for FC and TC combined, and Prasoon et al., (2013) reported 82.5% for TC. It is important to note that these studies did not explore the effect of their automatic segmentation on the mechanical response of the FE model generated from the segmentation. We noticed some abnormal smoothing, in the form of flat patches on the surface of cartilage, when applying previous automated segmentation methods. These poor local segmentations significantly affected local and global stress/strain responses, despite not affecting overall DSC or other metrics as described in Table 1. This study aimed to bridge this gap by integrating an automated segmentation and smoothing process in the development of knee cartilage geometries and evaluating the performance of the proposed method not only based on the quality of the generated surfaces but also through its impact on predicting mechanical responses in a FE study of 9 MRIs. This is to ensure that the discrepancy between automated segmentation and manual segmentation did not affect the prediction of mechanical responses.

The evaluation of the mechanical parameters at 20% and 80% of the stance phase, corresponding to the peaks of the loading condition (Figure 5), revealed a strong agreement in the distribution of the parameters between the manual and automated FE models. Figure 6 provides quantitative comparisons over the entire stance phase. Despite some discrepancies, no significant differences (p-value > 0.05) were observed (Figure 4). For instance, the largest error in fluid pressure was at 80% stance phase, which reflected in a higher t-value in SPM compared to other parameters (Figure 4). However, this difference was not significant as the respective values were well below the t-critical threshold.

The present study has several limitations. The meniscus and cartilage contacts were not incorporated in the FE modeling of knee cartilage, as highlighted in prior research (Danso et al., 2015; Simkheada et al., 2022). While the meniscus and cartilage contacts are crucial in knee biomechanics, their inclusion would substantially increase computational time for FE modeling. Considering the focus of our research, the decision to exclude these elements does not compromise the validity of our findings (Myller et al., 2020). Despite the fact that we selected MRIs with KL grades ranging from 0 to 4 in our study, the relatively small sample size may limit the generalizability of our findings. Future investigations may aim for larger sample sizes and explore automatic meshing techniques (Baldwin et al., 2010; Jahangir et al., 2022; Rodriguez-Vila et al., 2017). Additionally, the ligament spring elements did not account for the wrapping effect of ligaments, which might influence FE outcomes (Galbusera et al., 2014). However, since both models employed the same simplifications, the absence of ligament wrapping is unlikely to affect the interpretation of the results in our study. Another limitation was spatial resolution. Mechanical responses were obtained at element nodes or Gaussian points. Generating an identical mesh for the manual and automated FE models may not be feasible due to geometric differences. Consequently, the location of corresponding points used for comparison may vary slightly between models.

In conclusion, the integration of Swin UNETR segmentation model and the proposed automated filtering process demonstrated remarkable effectiveness in the simulation of knee cartilage. By leveraging the strengths of both Swin UNETR and automated filtering, this method generated appropriate shapes and geometries for FE models. We have made our automated segmentation models publicly available (https://data.mendeley.com/datasets/dc832g7j5m/1), aiming to advance biomechanical modeling and medical image segmentation. We hope this tool will enable the biomedical community to easily develop efficient subject-specific knee joint models for simulation studies.

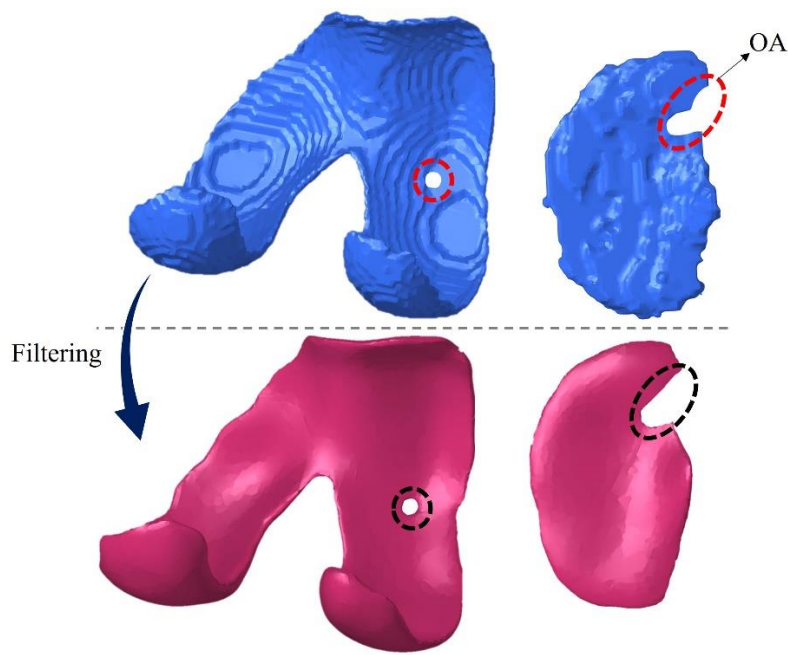

Figure 7. The proposed automated filtering preserved areas associated with OA while providing improved surface quality for FE modeling. Blue is the output of the Swin UNETR automated segmentation model. Red represents the model outcome after the automated smoothing process.

## Data availability:

Please refer to the https://data.mendeley.com/datasets/dc832g7j5m/1 for segmentation models. For OAI please refer to https://nda.nih.gov/oai/ (please email the website to request the images). FE models are available upon request to Reza Kakavand at reza.kakavand@ucalgary.ca.

**Acknowledgment:** The project was supported by the Natural Sciences and Engineering Research Council Canada (NSERC) Discovery grant [grant number 401610]; We would like to acknowledge and express our gratitude to Shirin Inanlou for her contribution to visual illustration and video preparation of supplementary material. Additionally, we would like to extend our appreciation to Alexander Tack for his assistance in the deep learning component. Their expertise and support greatly enhanced the quality of our research.

**Conflict of interest:** None

# *Supplementary Materials:*

# Swin UNETR segmentation with automated geometry filtering for biomechanical modeling of knee joint cartilage

Reza Kakavand[1], Peyman Tahghighi[1], Reza Ahmadi[1], W. Brent Edwards[1,2,3], Amin Komeili[1,2,3*]

[1]Department of Biomedical Engineering, Schulich School of Engineering, University of Calgary

[2]McCaig Institute for Bone and Joint Health, University of Calgary, Calgary, Canada.

[3]Human Performance Laboratory, Faculty of Kinesiology, University of Calgary, Calgary, Canada

**Corresponding author: Amin Komeili**

**\* [amin.komeili@ucalgary.ca](mailto:amin.komeili@ucalgary.ca)**

**Address: CCIT216, 2500 University Drive NW, Calgary, AB, T2N 1N4**

# 1 Methods

## 1.1 Computational modeling

**Constitutive equation:**

The mechanical response of matrix and fibril stresses were predicted using energy functions as described in Eq. (S 1):

$$\sigma = -pI + \emptyset_0 J^{-1} F \left( 2 \frac{\partial W_0(C)}{\partial C} \right) F^T + \emptyset_1 J^{-1} F \left( 2 \frac{\partial W_{1i}(C)}{\partial C} + 2 \frac{\partial \bar{W}_{1a}(\bar{C})}{\partial C} \right) F^T \quad \text{(S 1)}$$

where $J$ is the determinant of $F$ (deformation gradient matrix), $\bar{C}$ is the distortional component of the right Cauchy-Green deformation tensor, $C$. $W_0$ and $W_{1i}$ are the Holmes-Mow (Holmes & Mow, 1990) elastic strain energy potential of isotropic matrix and collagen fibrils, respectively, defined as:

$$W_{HM}(C) = \alpha_0 \frac{\exp[\alpha_1(I_1(C) - 3) + \alpha_2(I_2(C) - 3)]}{(I_3(C))^\beta} \quad \text{(S 2)}$$

Where $\alpha_0$, $\alpha_1$, $\alpha_2$ and $\beta$ are the material constants. $\bar{W}_{1a}$ was expressed as a function of the distortional component of $C$ in Eq. (S 2), because the collagen fibrils were assumed incompressible (Federico & Gasser, 2010):

$$\bar{W}_{1a}(\bar{C}) = \int_{S_X^2} \psi(\vec{M}) \times \frac{1}{2} c_{1b} \left[ \bar{I}_4 \left( \bar{C}, A(\vec{M}) \right) - 1 \right]^2 dS \quad \text{(S 3)}$$

where $c_{1b}$ is a material constant, $\bar{I}_4$ is the fourth invariant of $\bar{C}$, $A(\vec{M}) = \vec{M} \otimes \vec{M}$ is a structure tensor that is a function of fibrils direction in the reference configuration ($\vec{M}$). The $\psi(\vec{M})$ is a probability distribution density function that gives the probability of finding a fibril aligned with the direction $\vec{M}$:

$$\psi(\vec{M}) = \rho(\Theta) = \frac{1}{\pi} \sqrt{\frac{b}{2\pi}} \frac{\exp[b(\cos 2\Theta) + 1]}{erfi(\sqrt{2b})} \quad \text{(S 4)}$$

Where $\Theta$ is the co-latitude in polar coordinates, *erfi(x)* is the imaginary error function and the direction of the fibrils are controlled by the parameter b; negative and positive values of b produce parallel and perpendicular to surface fibrils orientation, while b=0 generates equally distribute fibrils over a sphere, i.e. random fibers distribution. Table 1.SM provides a description of the biphasic model and associated material constants.

**Material constants**

Table 1.SM illustrates the material parameters that were taken from Komeili et al. (2020) .

Table 1.SM. Material constants of the model.

| Material properties | Collagen Fibril | | ECM | |
|---|---|---|---|---|
| | SZ | DZ | SZ | DZ |
| $E$ (MPa)† | 10 | 15 | 2.5 | 3.8 |
| $v$ † | 0.3 | 0.3 | 0.1 | 0.1 |
| $\alpha_0$† | 3.4 | 5.1 | 0.6 | 1.0 |
| $\alpha_1$† | 0.1 | 0.1 | 0.8 | 0.8 |
| $\alpha_2$† | 0.4 | 0.4 | 0.1 | 0.1 |
| $c_{1b}$† (MPa) | 7.6 | 11.4 | – | – |
| $\beta$ † | 1.0 | 1.0 | 1.0 | 1.0 |
| $k$ † | – | – | 2.8 | 2.8 |
| $e_R$* † | – | – | 4.0 | 4.0 |
| b †† | 0 | 0 | 0 | 0 |
| Thickness‡ | 0.12h | 0.62h | 0.12h | 0.62h |

\* Void ratio (fluid / solid volume)

SZ: Superficial Zone

DZ: Deep Zone

h: Cartilage thickness

† (Pajerski, 2010)

†† (Komeili et al., 2020)

‡ (Julkunen et al., 2007)

## 1.2 Swin UNETR random 5-fold cross-validation

To assess the effectiveness of the Swin UNETR model, a random 5-fold cross-validation method was employed, with each fold comprising 405 MRIs for training and 102 MRIs for testing.

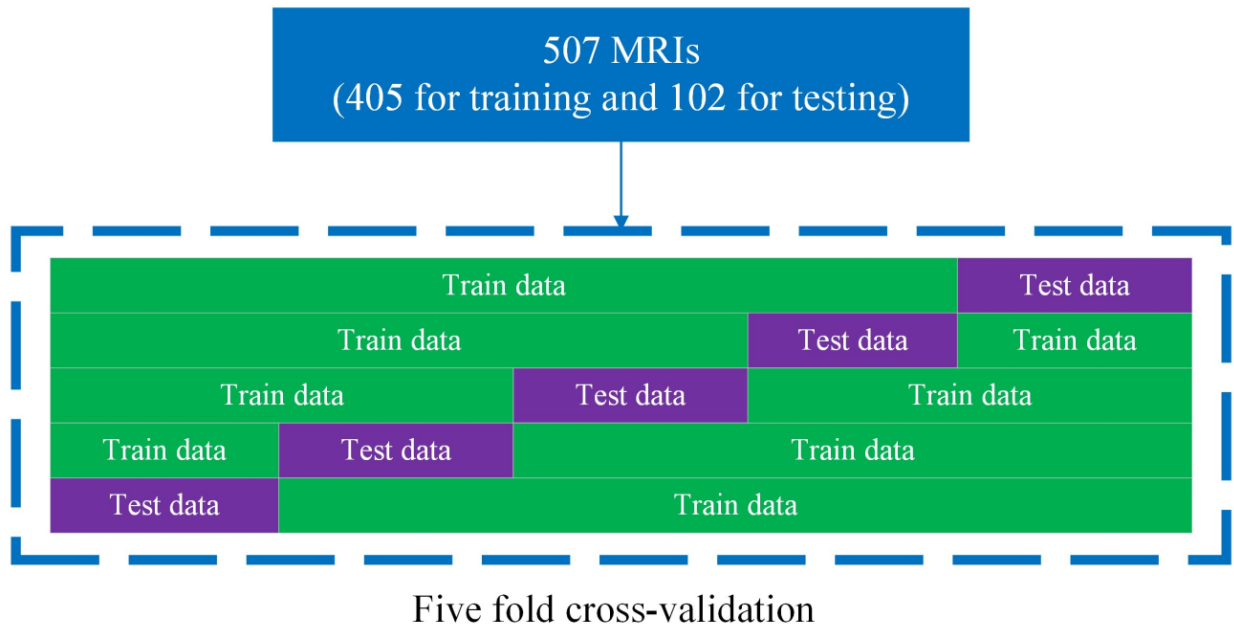

Figure 1.SM Five-fold cross-validation was used for training Swin UNETR models in our study.

## 1.3 Automated filtering

**Mesh repairing:**

The definition of each function used in the mesh repairing is as below:

- merging close vertices: "it merges all the vertices that are nearer than the specified threshold. It is the unification of duplicated vertices but with some tolerance". (Muntoni & Cignoni, 2021)
- merging close per-wedge texture coords: "it merges per-wedge texture coords that are very close. Used to correct apparent texture seams that can arise from numerical approximations when saving in ascii formats." (Muntoni & Cignoni, 2021)

- removing duplicate faces: "it deletes all the duplicate faces. Two faces are considered equal if they are composed by the same set of vertices, regardless of the order of the vertices." (Muntoni & Cignoni, 2021)
- removing duplicate vertices: "it checks for every vertex on the mesh: if there are two vertices with same coordinates, they are merged into a single one." (Muntoni & Cignoni, 2021)
- removing folded faces: "it deletes all the single folded faces. A face is considered folded if its normal is opposite to all the adjacent faces. It is removed by flipping it against the face f adjacent along the edge e such that the vertex opposite to e fall inside f." (Muntoni & Cignoni, 2021)
- removing null faces: "it deletes faces with area equal to zero". (Muntoni & Cignoni, 2021)
- removing floating faces: "it deletes isolated connected components (floating faces) whose diameter is smaller than the specified constant". (Muntoni & Cignoni, 2021)
- removing t-vertices: it arises when three or more aligned vertices exist. This function "deletes t-vertices from the mesh by edge collapse (collapsing the shortest of the incident edges) or edge flip (flipping the opposite edge on the degenerate face if the triangulation quality improves)." (Muntoni & Cignoni, 2021)
- removing unreferenced vertices: "it checks for every vertex on the mesh: if it is NOT referenced by a face, removes it" (Muntoni & Cignoni, 2021)
- removing non-manifold edges: it removes non-manifold edges by removing faces (for each non-manifold edge it iteratively deletes the smallest area face until it becomes 2-Manifold) or by splitting vertices (each non manifold edges chain will become a border). (Muntoni & Cignoni, 2021)
- removing non-manifold vertices: it splits non-manifold vertices until it becomes 2-Manifold. (Muntoni & Cignoni, 2021)

## 1.4 Knee cartilage mesh

The knee cartilage mesh was defined using hexahedral pore pressure elements. Figure 2.SM show the steps taken to generate the mesh for FC in HyperMesh 2019 (Altair Inc, Santa Ana, CA).

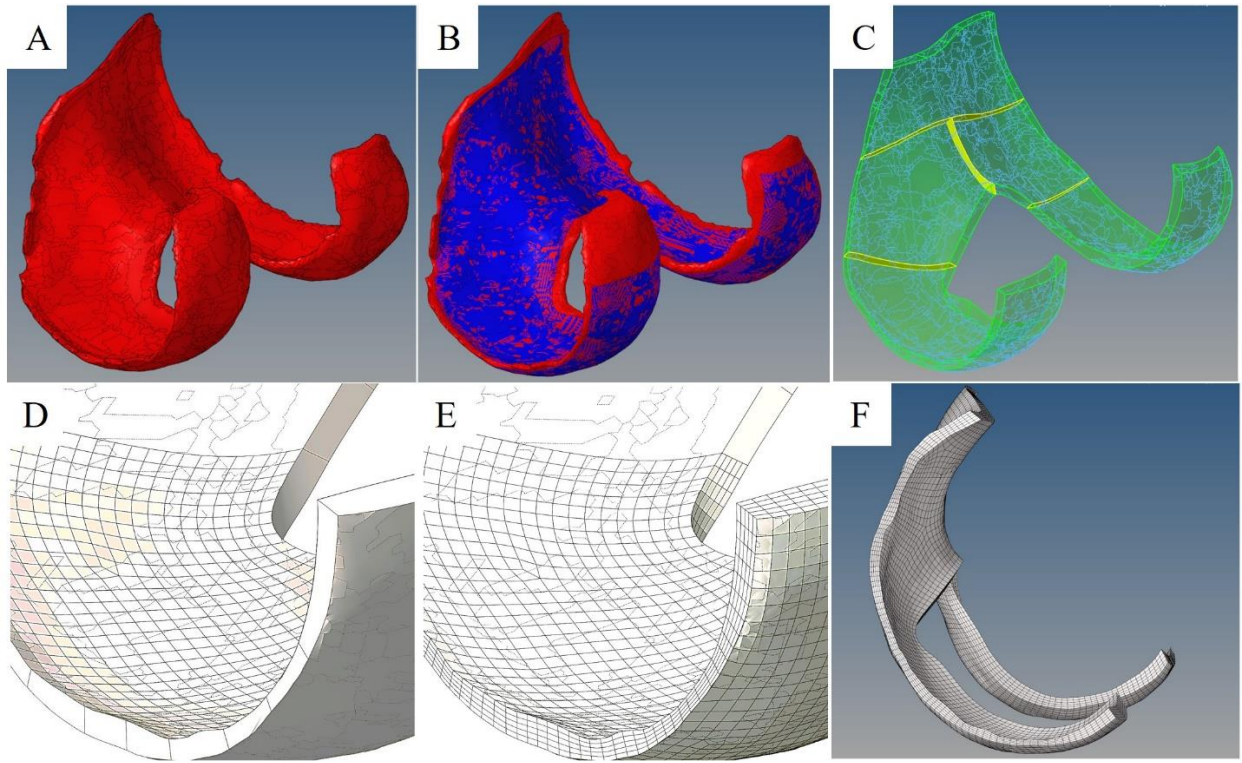

Figure 2.SM. Mesh generation: (A) 2D triangular elements of cartilage surfaces were converted to an enclosed volume and then were converted to a solid. (B) The jagged boundary of the cartilage perimeter was trimmed to facilitate 3D mesh generation. (C) The geometry was partitioned into different sections, giving more control over the mesh. (D) For each partition, the face that is shared between the cartilage and calcified zone was meshed. (E) The 2D mesh of the calcified zone was mapped to the articular surface and a depth-wise element size was created. (F) A fully meshed cartilage model was illustrated.

# 2 Results

## 2.1 Swin UNETR

In Figure 3.SM, we present the results of the MRI segmentation comparison between the manual segmentation and the Swin UNETR segmentation method. The purpose of this evaluation is to visually assess the performance of the Swin UNETR model in segmenting MRI and to compare it against the manual segmentation. The green contour overlaid on the Swin UNETR segmentation indicates the alignment with the outline obtained from manual segmentation.

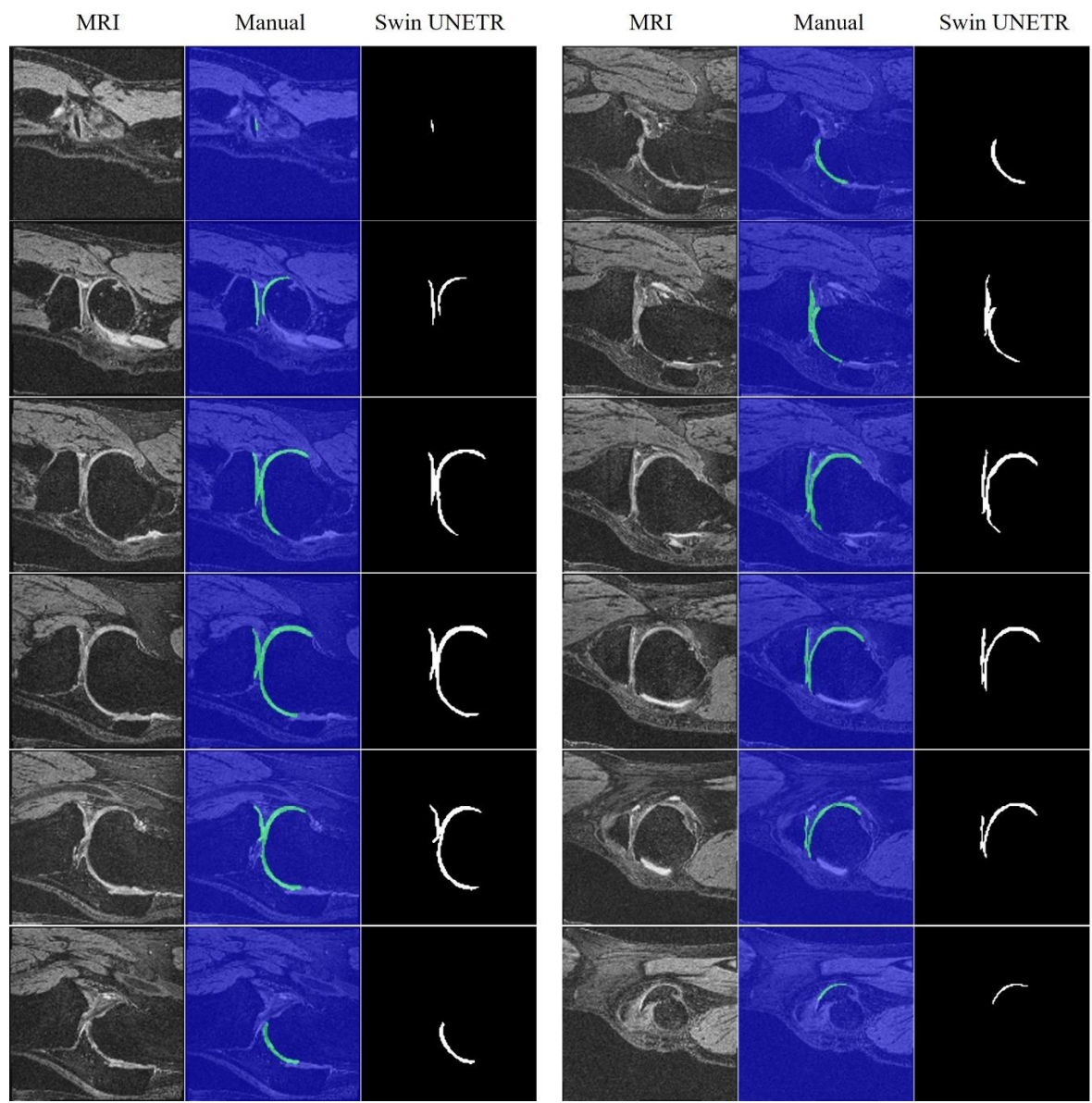

Figure 3.SM. Example of MRI segmentation results comparison: manual segmentation and Swin UNETR segmentation. The green contour on Swin UNETR segmentation is the outline from the manual segmentation.

## Mechanical Responses

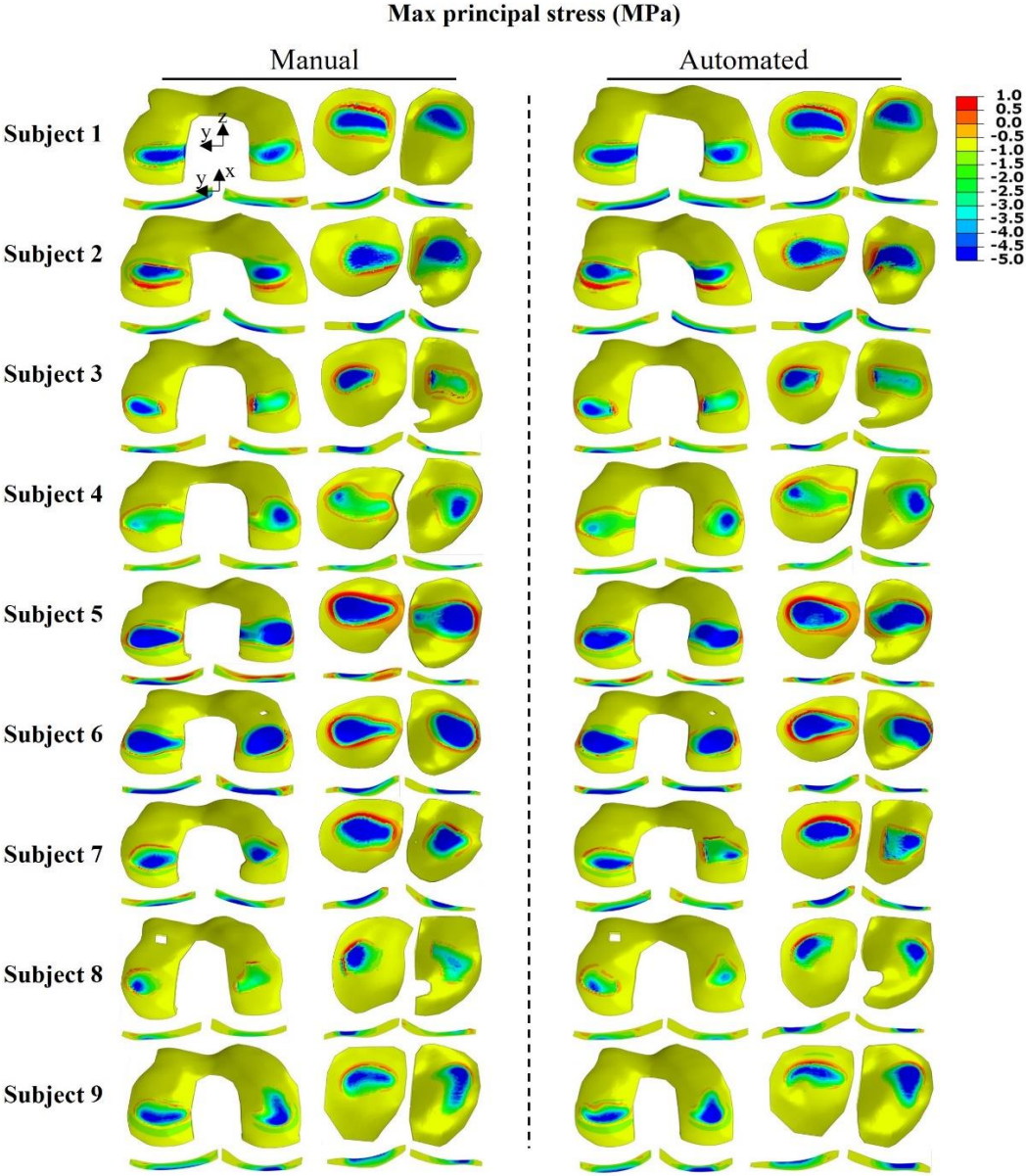

Figure 4.SM. The distribution of maximum principal stress over the surface and along the thickness of nine cartilage models at 20% of the stance phase. The depth-wise illustration was from the cross-section where the peak value occurred.

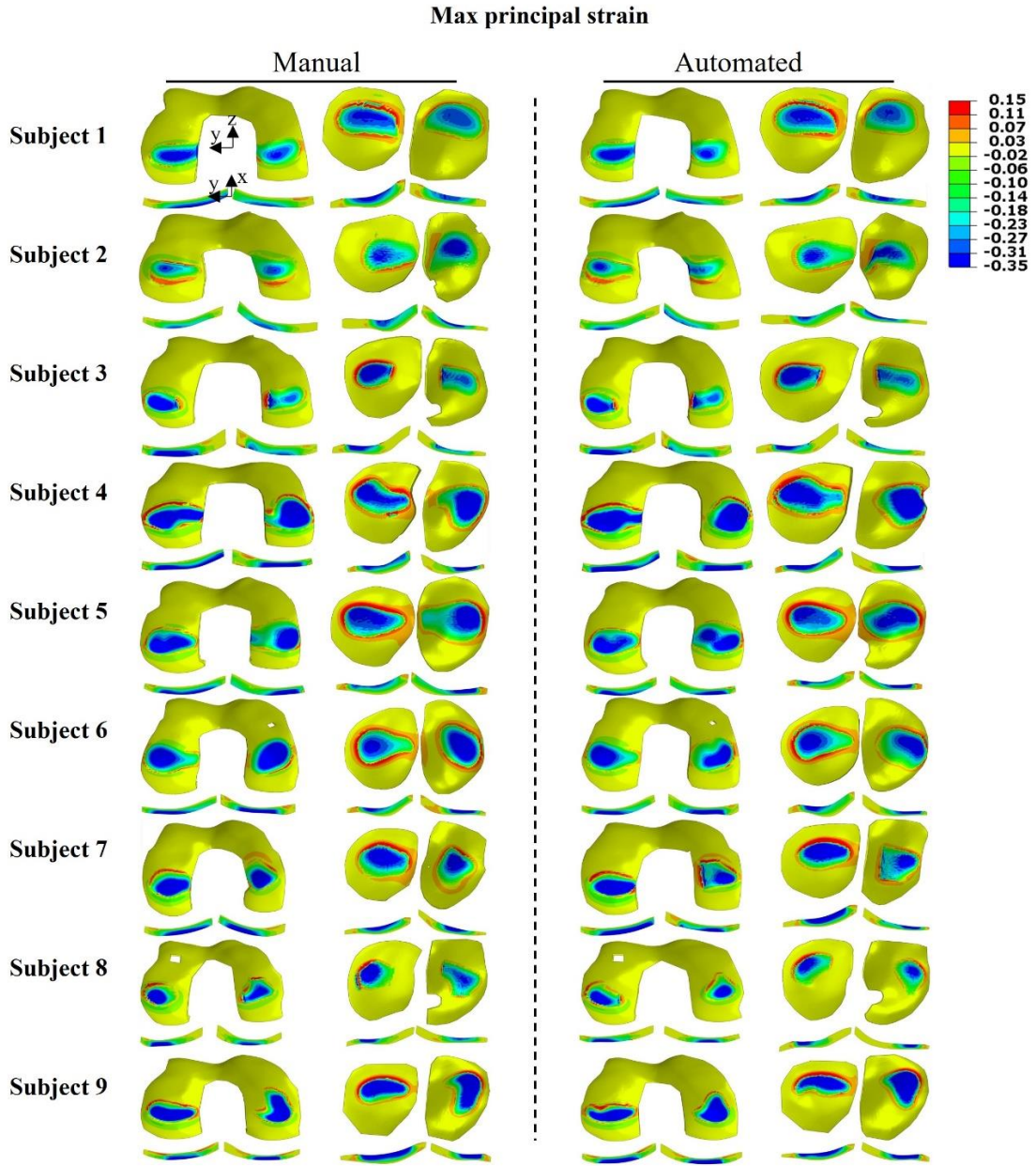

Figure 5.SM. The distribution of maximum principal strain over the surface and along the thickness of nine cartilage models at 20% of the stance phase. The depth-wise illustration was from the cross-section where the peak value occurred.

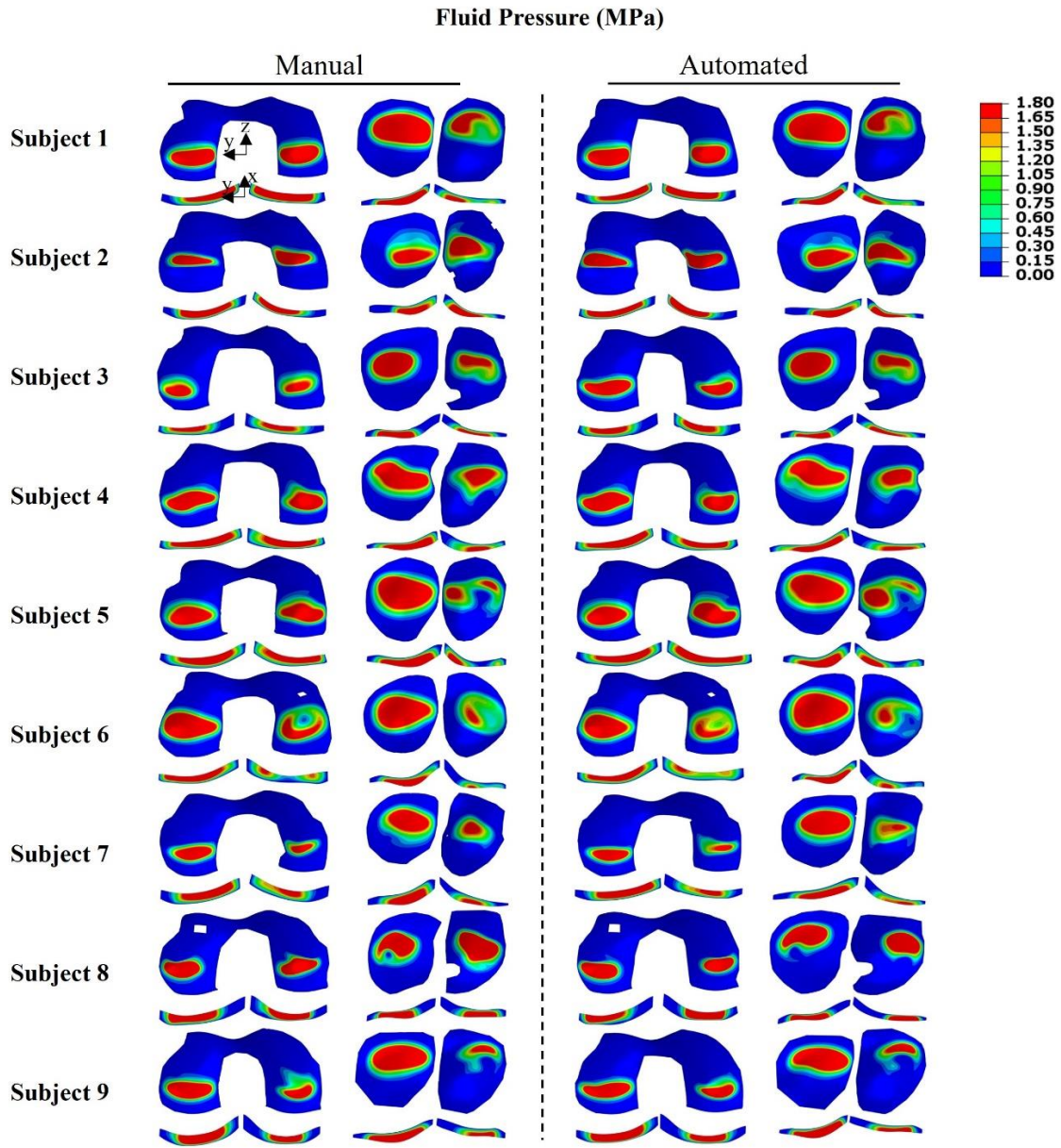

Figure 6.SM. The distribution of fluid pressure in the superficial zone and along the thickness of nine cartilage models at 20% of the stance phase. The depth-wise illustration was from the cross-section where the peak value occurred.

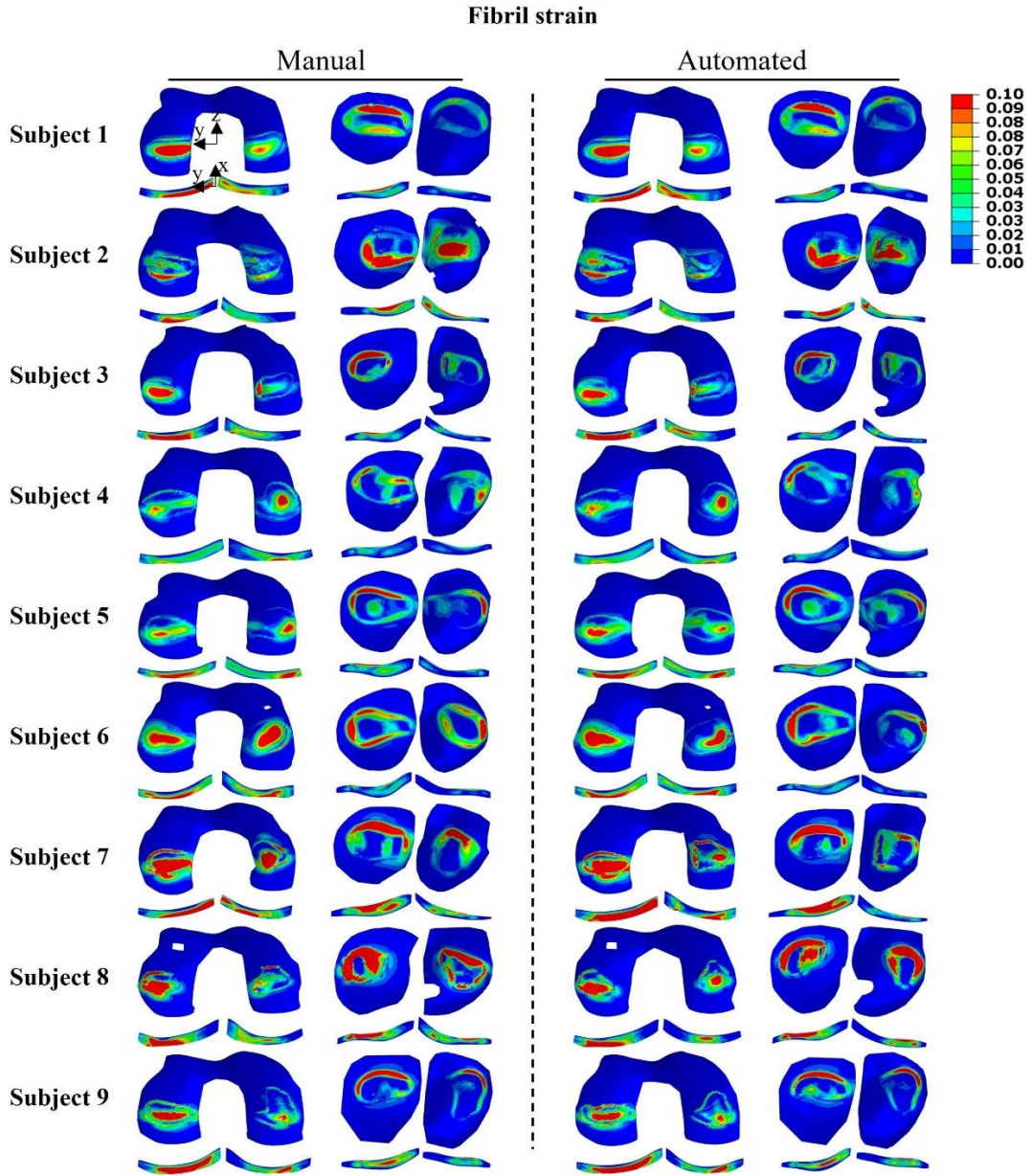

Figure 7.SM. The distribution of fibril strain in over the surface and along the thickness of nine cartilage models at 20% of the stance phase. The depth-wise illustration was from the cross-section where the peak value occurred.

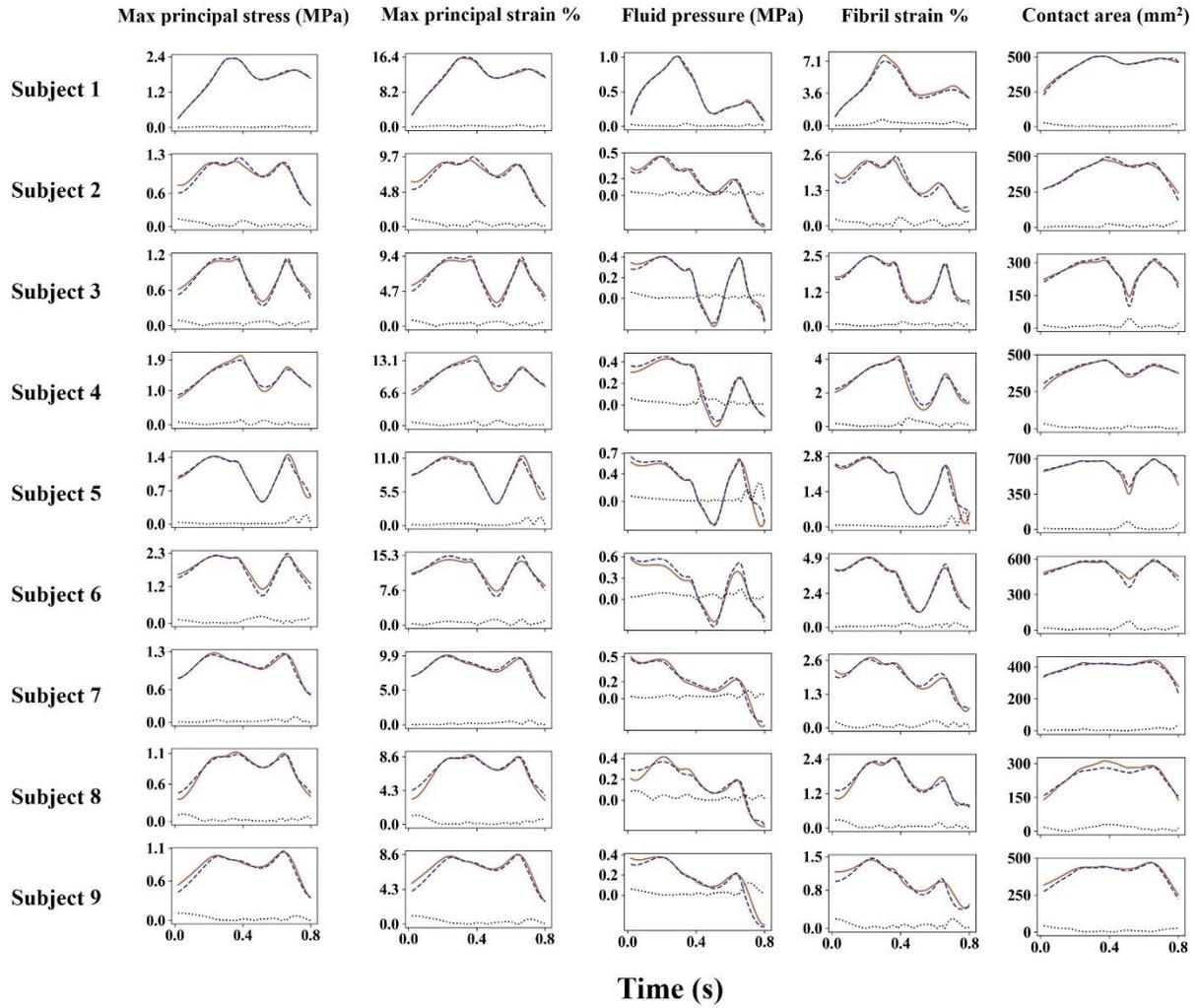

Figure 8.SM. The **average** values of the mechanical parameters in the **superficial zone** for all cartilage models. The solid and dashed lines represent the manual and automated models, respectively. The dotted line is the absolute difference between the two models. Values were calculated from the contact region.

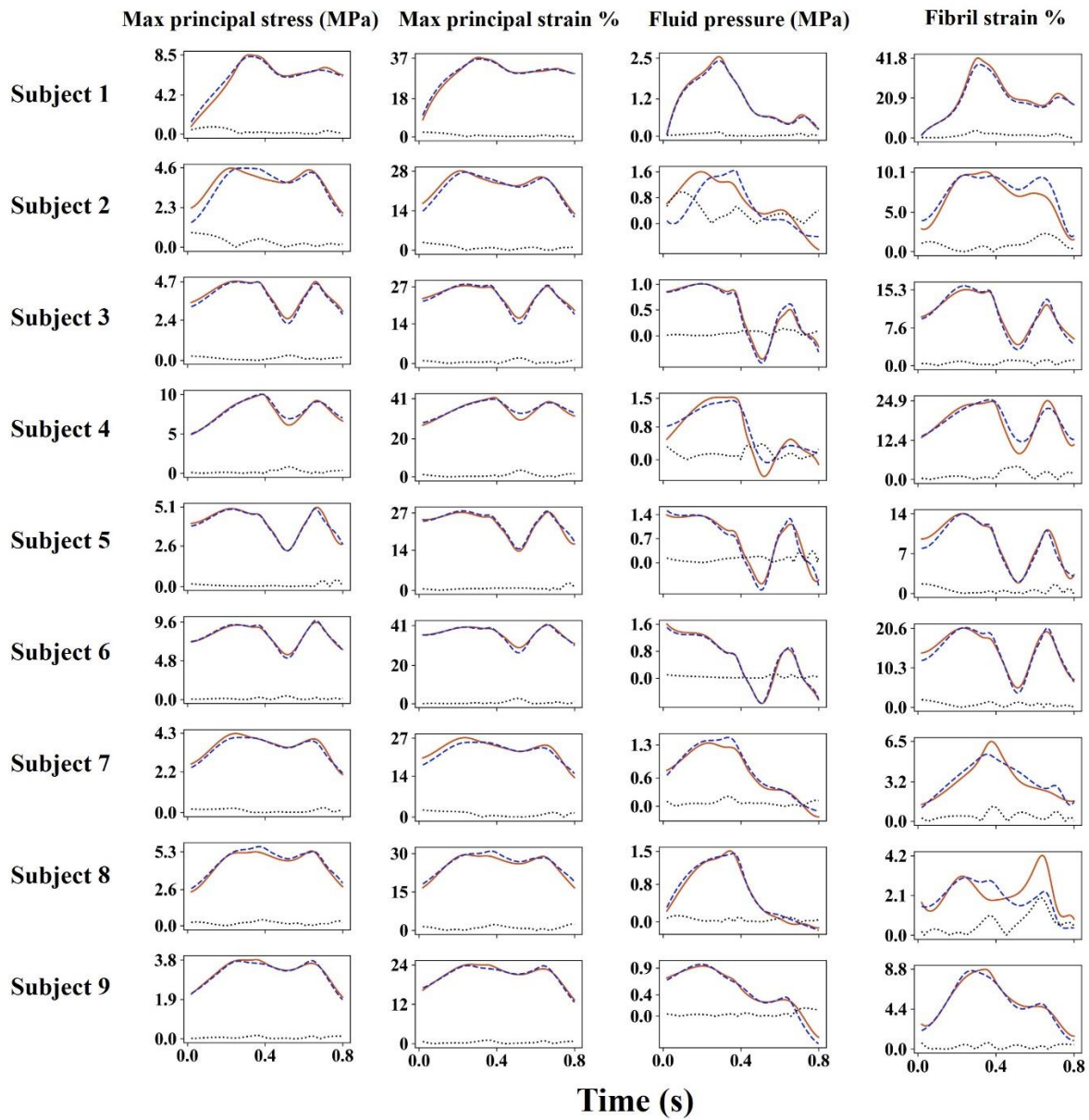

Figure 9.SM. The **peak** values of the mechanical parameters in the **superficial zone** for all cartilage models. The solid and dashed lines represent the manual and automated models, respectively. The dotted line is the absolute difference between the two models. Values were calculated from the contact region.

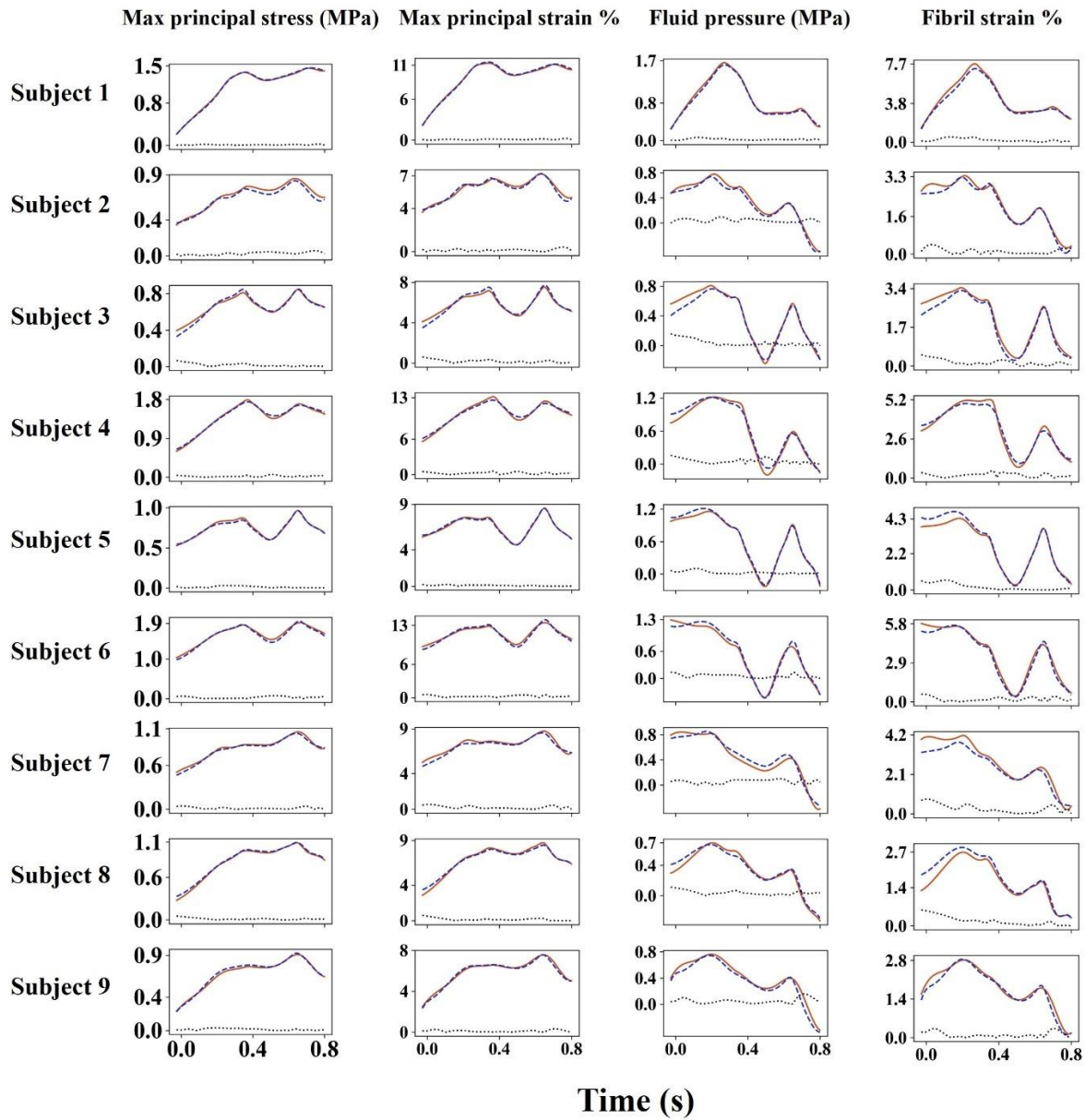

Figure 10.SM. The **average** values of the mechanical parameters in the **deep zone** for all cartilage models. The solid and dashed lines represent the manual and automated models, respectively. The dotted line is the absolute difference between the two models. The contact region of the superficial zone was projected into the deep zone for calculating the parameters.

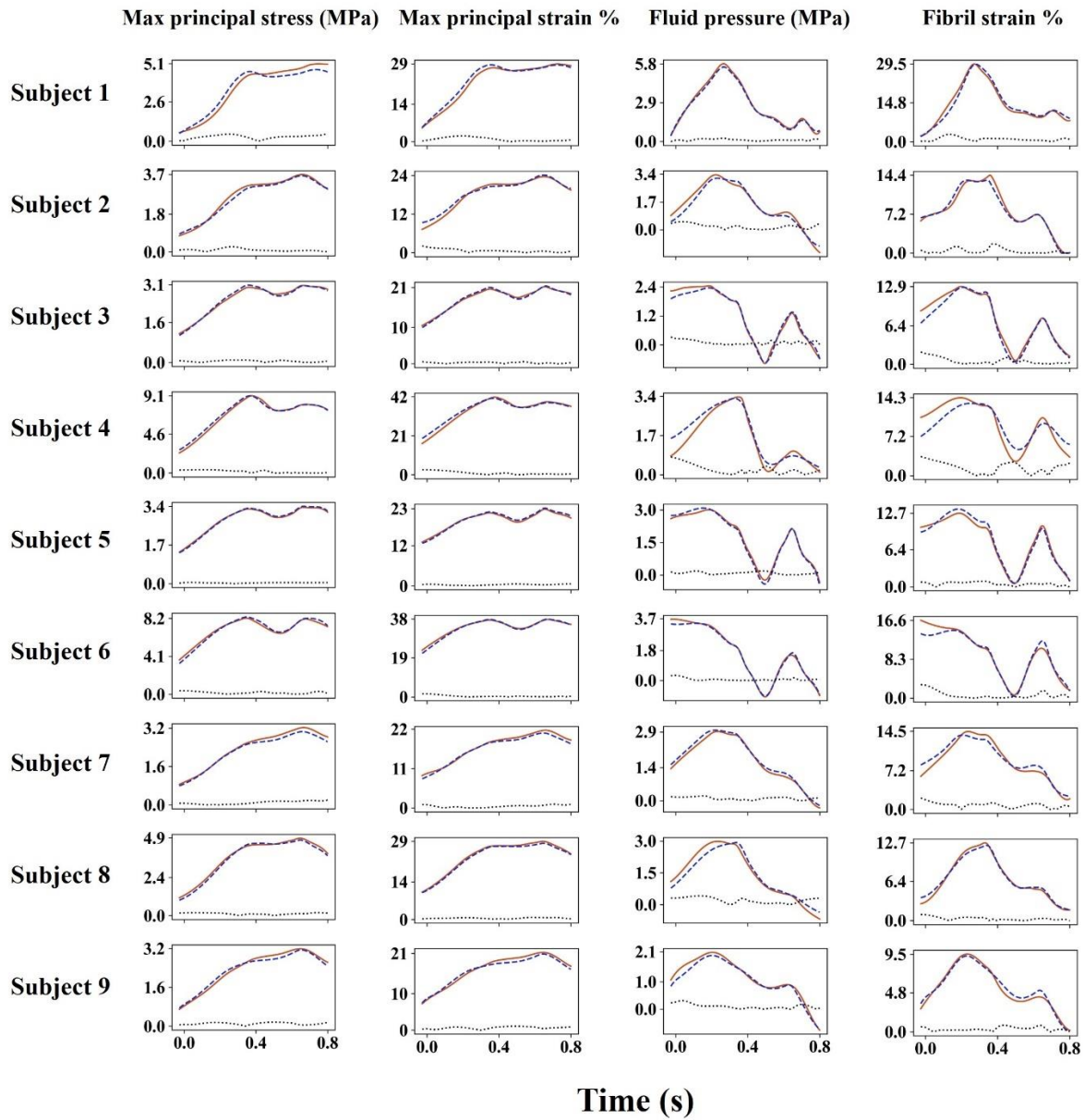

Figure 11.SM. The **peak** values of the mechanical parameters in the **deep zone** for all cartilage models. The solid and dashed lines represent the manual and automated models, respectively. The dotted line is the absolute difference between the two models. The contact region of the superficial zone was projected into the deep zone for calculating the parameters.